\documentclass[aps,preprint,showkeys,tightenlines]{revtex4}

\usepackage{bm}

\begin{document}

\title{Symmetry limit properties of decay amplitudes with mirror
matter admixtures}

\author{G.~S\'anchez-Col\'on}

\email[]{gsanchez@mda.cinvestav.mx}

\affiliation{
Departamento de F\'{\i}sica Aplicada.\\
Centro de Investigaci\'on y de Estudios Avanzados del IPN.\\
Unidad Merida.\\
A.P. 73, Cordemex. \\
M\'erida, Yucat\'an, 97310. MEXICO.
}

\author{A.~Garc\'{\i}a}

\affiliation{
Departamento de F\'{\i}sica.\\
Centro de Investigaci\'on y de Estudios Avanzados del IPN.\\
A.P. 14-740.\\
M\'exico, D.F., 07000. MEXICO.
}

\date{\today}

\begin{abstract}

We extend our previous analysis on the symmetry
limit properties of non-leptonic and weak radiative decay
amplitudes of hyperons in a scheme of mirror matter admixtures
in physical hadrons to include the two-body non-leptonic decays
of $\Omega^-$ and the two photon and two pion decays
of kaons. We show that the so-called parity-conserving
amplitudes predicted for all the decays vanish in the strong
flavor SU(3) symmetry limit. We also establish the specific
conditions under which the corresponding so-called
parity-violating amplitudes vanish in the same limit.

\end{abstract}



\maketitle

\section{Introduction}
\label{introduction}

Two interesting paths implement the violations of the conservation laws of
parity and strong flavors. One is via the perturbative intervention of the
$W^{\pm}$ and $Z^0$ vector mesons, as in the Standard Model (SM). The other
one is via non-perturbative superpositions of different parity and flavor
eigenstates in the physical states, eigenstates of the Hamiltonian and
mass operators, which arise from the fact that these operators do not
commute with the parity and flavor operators. This second path cannot be
implemented in the minimal SM. However, it emerges quite naturally in
extensions of the SM and, thus, may indicate the existence of some forms
of new physics. It is therefore important to see if and how this second
form of parity and flavor violation is relevant in low-energy physics.
Through a systematic study we have established that, indeed, it can give
significant contributions to the enhancement phenomenon observed in
non-leptonic, weak radiative, and some rare decay modes of strange baryons
and
mesons~\cite{apriori,universality,detailed,omegasymlim,kl2g,ks2g,kpipi}.
This we have done with a phenomenological model we have referred
to as manifest mirror matter admixtures in ordinary hadrons.

The transition operators in the amplitudes of these decays are the
parity and flavor conserving strong $H_{\rm st}$ and electromagnetic
$H_{\rm em}$ parts of the exact Hamiltonian $H$. Although both of
them break the strong flavor symmetries, we have considered in
Ref.~\cite{symmetry} some of  the consequences of assuming that they
do not, i.e., the consequences on the decay amplitudes of the
symmetry limit behavior of these two operators. Only the so-called
parity-conserving amplitudes of non-leptonic and weak radiative
decays of hyperons (NLDH and WRDH) were studied in
Ref.~\cite{symmetry}.

In this paper we extend our previous analysis of
Ref.~\cite{symmetry} to include the two-body non-leptonic decays
of the $\Omega^-$, the two photon decays of $K_{\rm L}$ and
$K_{\rm S}$, and the two pion decays of $K$ mesons. We find
that in the manifest mirror matter admixtures scheme, the
so-called parity-conserving amplitudes in all these decays
vanish automatically in the strong flavor SU(3) symmetry limit.
This extends the result obtained in Ref.~\cite{symmetry} for
NLDH and WRDH. We also find in the present analysis the specific
and systematic conditions under which the same conclusion is
valid for the corresponding so-called parity-violating decay
amplitudes. In Section~\ref{mixings} we reproduce the
expressions for the physical (mass eigenstates) hadrons with
mirror matter admixtures obtained in our previous work,
Ref.~\cite{apriori}. In Secs.~\ref{nldh} and \ref{onl} we study
NLDH and the two-body non-leptonic decays of $\Omega^-$,
respectively. Section~\ref{wrdh} is dedicated to WRDH. We
complement our analysis with the inclusion of the $2\gamma$
decays of $K_{\rm L}$ and $K_{\rm S}$ in Sec.~\ref{ksl} and of
the two pion decays of $K$ mesons in Sec.~\ref{ktopipi}. We
present our conclusions in Sec.~\ref{conclusions}.

\section{Mirror matter admixtures in physical hadrons}
\label{mixings}

Let us start with the expressions for the physical hadrons we shall
use in terms of the mirror matter admixtures~\cite{apriori}. The
SU(3) octets of spin-0 mesons and spin-1/2 baryons are given by

\[
K^+_{\rm ph} =
K^+_{\rm p} - \sigma \pi^+_{\rm p} - \delta' \pi^+_{\rm s}
+ \cdots
,
\]

\[
K^0_{\rm ph} = K^0_{\rm p} + \sigma (\frac{1}{\sqrt{2}} \pi^0_{\rm p}
+ \sqrt{\frac{3}{2}} \eta_{\rm 8p}) + \delta (\sqrt{\frac{2}{3}}
\eta_{\rm 8s} + \frac{1}{\sqrt{3}} \eta_{\rm 1s}) + \delta'
(\frac{1}{\sqrt{2}} \pi^0_{\rm s} + \frac{1}{\sqrt{6}}
\eta_{\rm 8s} - \frac{1}{\sqrt{3}} \eta_{\rm 1s})
+ \cdots
,
\]
 
\[
\pi^+_{\rm ph} =
\pi^+_{\rm p} + \sigma K^+_{\rm p} - \delta K^+_{\rm s}
+ \cdots
,
\]

\begin{equation}
\pi^0_{\rm ph} =
\pi^0_{\rm p} -
\sigma \frac{1}{\sqrt{2}} ( K^0_{\rm p} + \bar{K}^0_{\rm p} ) +
\delta \frac{1}{\sqrt{2}} ( K^0_{\rm s} - \bar{K}^0_{\rm s} )
+ \cdots
,
\label{mph}
\end{equation}
 
\[
\pi^-_{\rm ph} =
\pi^-_{\rm p} + \sigma K^-_{\rm p} + \delta K^-_{\rm s}
+ \cdots
,
\]
 
\[
\bar{K}^0_{\rm ph} =
\bar{K}^0_{\rm p} + \sigma (\frac{1}{\sqrt{2}} \pi^0_{\rm p}
+ \sqrt{\frac{3}{2}} \eta_{\rm 8p})
- \delta (\sqrt{\frac{2}{3}} \eta_{\rm 8s} + \frac{1}{\sqrt{3}}
\eta_{\rm 1s}) - \delta' (\frac{1}{\sqrt{2}} \pi^0_{\rm s} +
\frac{1}{\sqrt{6}} \eta_{\rm 8s} - \frac{1}{\sqrt{3}} \eta_{\rm 1s})
+ \cdots
,
\]

\[
K^-_{\rm ph} =
K^-_{\rm p} - \sigma \pi^-_{\rm p} + \delta' \pi^-_{\rm s}
+ \cdots
,
\]
 
\[
\eta_{\rm 8ph} =
\eta_{\rm 8p} -
\sigma \sqrt{\frac{3}{2}}( K^0_{\rm p} + \bar{K}^0_{\rm p} ) +
(\delta + 2\delta') \frac{1}{\sqrt{6}} ( K^0_{\rm s} -
\bar{K}^0_{\rm s} )
+ \cdots
,
\]
 
\[
\eta_{\rm 1ph} =
\eta_{\rm 1p} -
(\delta - \delta') \frac{1}{\sqrt{3}} ( K^0_{\rm s} -
\bar{K}^0_{\rm s} ) + \cdots
,
\]

\noindent
and,
 
\[
p_{\rm ph} =
p_{\rm s} - \sigma \Sigma^+_{\rm s} - \delta \Sigma^+_{\rm p}
+ \cdots
,
\]
 
\[
n_{\rm ph} = n_{\rm s} + \sigma (\frac{1}{\sqrt{2}} \Sigma^0_{\rm s}
+ \sqrt{\frac{3}{2}} \Lambda_{\rm s}) + \delta (\frac{1}{\sqrt{2}}
\Sigma^0_{\rm p} + \sqrt{\frac{3}{2}} \Lambda_{\rm p})
+ \cdots
,
\]

\[
\Sigma^+_{\rm ph} =
\Sigma^+_{\rm s} + \sigma p_{\rm s} - \delta' p_{\rm p}
+ \cdots
,
\]
 
\begin{equation}
\Sigma^0_{\rm ph} = \Sigma^0_{\rm s} + \sigma \frac{1}{\sqrt{2}}
(\Xi^0_{\rm s}- n_{\rm s} ) + \delta \frac{1}{\sqrt{2}}
\Xi^0_{\rm p} + \delta'\frac{1}{\sqrt{2}} n_{\rm p}
+ \cdots
,
\label{bph}
\end{equation}
 
\[
\Sigma^-_{\rm ph} = \Sigma^-_{\rm s} + \sigma \Xi^-_{\rm s} + \delta
\Xi^-_{\rm 0p}
+ \cdots
,
\]

\[
\Lambda_{\rm ph} =
\Lambda_{\rm s} +
\sigma \sqrt{\frac{3}{2}} (\Xi^0_{\rm s}- n_{\rm s}) +
\delta \sqrt{\frac{3}{2}} \Xi^0_{\rm p} +
\delta' \sqrt{\frac{3}{2}} n_{\rm p}
+ \cdots
,
\]
 
\[ \Xi^0_{\rm ph} = \Xi^0_{\rm s} - \sigma (\frac{1}{\sqrt{2}}
\Sigma^0_{\rm s} + \sqrt{\frac{3}{2}} \Lambda_{\rm s}) +
\delta' (\frac{1}{\sqrt{2}} \Sigma^0_{\rm p} +
\sqrt{\frac{3}{2}} \Lambda_{\rm p})
+ \cdots
,
\]

\[
\Xi^-_{\rm ph} =
\Xi^-_{\rm s} - \sigma \Sigma^-_{\rm s} + \delta' \Sigma^-_{\rm p}
+ \cdots
.
\]

\noindent
The dots stand for other mixings (with strong flavors other than
strangeness) that will not be relevant here. The subindices s and p
refer to positive and negative parity eigenstates, respectively.
$\delta$, $\delta'$, and $\sigma$ are the mixing angles. Each
physical hadron is the mass eigenstate observed in experiment
and our phase conventions are those of Ref.~\cite{gibson}.

\section{Two-body Non-leptonic Decay Amplitudes of Hyperons}
\label{nldh}

As mentioned in Sec.~\ref{introduction}, mirror matter mixings in hadrons
lead to NLDH via the parity and flavor conserving strong part $H_{\rm st}$
of the Hamiltonian. The transition amplitudes will be given by the matrix
elements $\langle M_{\rm ph}B'_{\rm ph}|H_{\rm st}|B_{\rm ph}\rangle$,
where $B_{\rm ph}$ and $B'_{\rm ph}$ are the initial and final hyperons
and $M_{\rm ph}$ is the emitted meson. Using the above mixings these
amplitudes will have the form $\bar{u}_{B'}(A-B\gamma_5)u_B$, where $u_B$
and $u_{B'}$ are four-component Dirac spinors and the amplitudes $A$ and
$B$ correspond to the parity-violating and the parity-conserving
amplitudes of the $W^{\pm}_{\mu}$ mediated NLDH, although with mirror
matter mixings these amplitudes are both actually parity and flavor
conserving. These amplitudes are given by~\cite{detailed}

\[
A_1
=
\delta'
\sqrt{\frac{3}{2}} g^{{}^{\rm p,sp}}_{{}_{n,p\pi^-}} +
\delta
(
g^{{}^{\rm s,ss}}_{{}_{\Lambda,pK^-}} - g^{{}^{\rm
s,pp}}_{{}_{\Lambda,\Sigma^+\pi^-}} ) , \]

\[
A_2
=
-\frac{1}{\sqrt{2}}
[
-\delta'
\sqrt{3}g^{{}^{\rm p,sp}}_{{}_{n,n\pi^0}} +
\delta
(
g^{{}^{\rm s,ss}}_{{}_{\Lambda,n\bar{K}^0}} -
\sqrt{3} g^{{}^{\rm s,pp}}_{{}_{\Lambda,\Lambda\pi^0}} -
g^{{}^{\rm s,pp}}_{{}_{\Lambda,\Sigma^0\pi^0}}
)
]
,
\]

\[
A_3
=
\delta
(
g^{{}^{\rm s,ss}}_{{}_{\Sigma^-,nK^-}} +
\sqrt{\frac{3}{2}} g^{{}^{\rm s,pp}}_{{}_{\Sigma^-,\Lambda\pi^-}} +
\frac{1}{\sqrt{2}} g^{{}^{\rm s,pp}}_{{}_{\Sigma^-,\Sigma^0\pi^-}}
)
,
\]

\begin{equation}
A_4
=
-\delta'
g^{{}^{\rm p,sp}}_{{}_{p,n\pi^+}} +
\delta
(
\sqrt{\frac{3}{2}} g^{{}^{\rm s,pp}}_{{}_{\Sigma^+,\Lambda\pi^+}} +
\frac{1}{\sqrt{2}} g^{{}^{\rm s,pp}}_{{}_{\Sigma^+,\Sigma^0\pi^+}}
)
,
\label{aes}
\end{equation}

\[
A_5
=
- \delta'
g^{{}^{\rm p,sp}}_{{}_{p,p\pi^0}} -
\delta
(
\frac{1}{\sqrt{2}} g^{{}^{\rm s,ss}}_{{}_{\Sigma^+,p\bar{K}^0}} +
g^{{}^{\rm s,pp}}_{{}_{\Sigma^+,\Sigma^+\pi^0}}
)
,
\]

\[
A_6
=
\delta'
g^{{}^{\rm p,sp}}_{{}_{\Sigma^-,\Lambda\pi^-}} +
\delta
(
g^{{}^{\rm s,ss}}_{{}_{\Xi^-,\Lambda K^-}} +
\sqrt{\frac{3}{2}} g^{{}^{\rm s,pp}}_{{}_{\Xi^-,\Xi^0\pi^-}}
)
,
\]

\[
A_7
=
\frac{1}{\sqrt{2}}
[
\delta'
(
\sqrt{3} g^{{}^{\rm p,sp}}_{{}_{\Lambda,\Lambda\pi^0}} +
g^{{}^{\rm p,sp}}_{{}_{\Sigma^0,\Lambda\pi^0}}
)
+
\delta
(
- g^{{}^{\rm s,ss}}_{{}_{\Xi^0,\Lambda\bar{K}^0}} +
\sqrt{3} g^{{}^{\rm s,pp}}_{{}_{\Xi^0,\Xi^0\pi^0}}
)
]
,
\]

\noindent
and

\[
B_1
=
\sigma
(
- \sqrt{\frac{3}{2}} g_{{}_{n,p\pi^-}} +
g_{{}_{\Lambda,pK^-}} - g_{{}_{\Lambda,\Sigma^+\pi^-}}
)
,
\]

\[
B_2
=
-\frac{1}{\sqrt{2}}
\sigma
(
\sqrt{3}g_{{}_{n,n\pi^0}} +
g_{{}_{\Lambda,n\bar{K}^0}} -
\sqrt{3} g_{{}_{\Lambda,\Lambda\pi^0}} -
g_{{}_{\Lambda,\Sigma^0\pi^0}}
)
,
\]

\[
B_3
=
\sigma
(
g_{{}_{\Sigma^-,nK^-}} +
\sqrt{\frac{3}{2}} g_{{}_{\Sigma^-,\Lambda\pi^-}} +
\frac{1}{\sqrt{2}} g_{{}_{\Sigma^-,\Sigma^0\pi^-}}
)
,
\]

\begin{equation}
B_4
=
\sigma
(
g_{{}_{p,n\pi^+}} +
\sqrt{\frac{3}{2}} g_{{}_{\Sigma^+,\Lambda\pi^+}} +
\frac{1}{\sqrt{2}} g_{{}_{\Sigma^+,\Sigma^0\pi^+}}
)
,
\label{bes}
\end{equation}

\[
B_5
=
\sigma
(
g_{{}_{p,p\pi^0}} -
\frac{1}{\sqrt{2}} g_{{}_{\Sigma^+,p\bar{K}^0}} -
g_{{}_{\Sigma^+,\Sigma^+\pi^0}}
)
,
\]

\[
B_6
=
\sigma
(
- g_{{}_{\Sigma^-,\Lambda\pi^-}} +
g_{{}_{\Xi^-,\Lambda K^-}} +
\sqrt{\frac{3}{2}} g_{{}_{\Xi^-,\Xi^0\pi^-}}
)
,
\]

\[
B_7
=
\frac{1}{\sqrt{2}}
\sigma
(
- \sqrt{3} g_{{}_{\Lambda,\Lambda\pi^0}} -
g_{{}_{\Sigma^0,\Lambda\pi^0}} -
g_{{}_{\Xi^0,\Lambda\bar{K}^0}} +
\sqrt{3} g_{{}_{\Xi^0,\Xi^0\pi^0}}
)
.
\]

\noindent
The subindices $1, \dots, 7$ correspond to $\Lambda\to p\pi^-$,
$\Lambda\to n\pi^0$, $\Sigma^-\to n\pi^-$, $\Sigma^+\to n\pi^+$,
$\Sigma^+\to p\pi^0$, $\Xi^-\to \Lambda\pi^-$, and $\Xi^0\to
\Lambda\pi^0$, respectively. The $g$-constants in these equations are
Yukawa coupling constants (YCC) defined by the matrix elements of
$H_{\rm st}$ between flavor and parity eigenstates, for example, by
$\langle M_{\rm p} B'_{\rm p} |H_{\rm st}|B_{\rm s}\rangle=g^{{}^{\rm
s,pp}}_{{}_{B,B'M}}$. The YCC in the $B$'s are the ordinary ones, while
the YCC in the $A$'s are new ones. In the latter, the upper indices serve
as a reminder of the parities of the parity eigenstates involved. We have
omitted the upper indices in the $g$'s of the $B$ amplitudes because the
states involved carry the normal intrinsic parities of hadrons. However,
because of our assumptions, all the ordinary and new YCC have common
properties.

The SU(2) symmetry limit of the YCC leads to the equalities

\[
g_{{}_{p,p\pi^0}}=-g_{{}_{n,n\pi^0}}=\frac{1}{\sqrt 2}g_{{}_{p,n\pi^+}}
=\frac{1}{\sqrt 2}g_{{}_{n,p\pi^-}},
\]

\[
g_{{}_{\Sigma^+,\Lambda\pi^+}}=g_{{}_{\Sigma^0,\Lambda\pi^0}}
=g_{{}_{\Sigma^-,\Lambda\pi^-}},
~~~~
g_{{}_{\Lambda,\Sigma^+\pi^-}}=g_{{}_{\Lambda,\Sigma^0\pi^0}},
\]

\begin{equation}
g_{{}_{\Sigma^+,\Sigma^+\pi^0}}=-g_{{}_{\Sigma^+,\Sigma^0\pi^+}}
=g_{{}_{\Sigma^-,\Sigma^0\pi^-}},
\label{su2limit}
\end{equation}

\[
g_{{}_{\Sigma^0,pK^-}}=\frac{1}{\sqrt 2}g_{{}_{\Sigma^-,nK^-}}
=\frac{1}{\sqrt 2}g_{{}_{\Sigma^+,p\bar K^0}},
\]

\[
g_{{}_{\Lambda,pK^-}}=g_{{}_{\Lambda,n\bar K^0}},
~~~~
g_{{}_{\Xi^0,\Xi^0\pi^0}}=\frac{1}{\sqrt 2}g_{{}_{\Xi^-,\Xi^0\pi^-}},
\]

\[
g_{{}_{\Xi^-,\Lambda K^-}}=-g_{{}_{\Xi^0,\Lambda \bar K^0}},
~~~~
g_{{}_{\Lambda,\Lambda \pi^0}}=0.
\]

\noindent
Similar relations are valid within each set of upper
indices, e.g.\ $g^{{}^{\rm p,sp}}_{{}_{p,p\pi^0}}= -g^{{}^{\rm
p,sp}}_{{}_{n,n\pi^0}}$, etc.\ when SU(2) symmetry is applied to
the new YCC. In the SU(3) limit one also has~\cite{gibson}

\[
g_{{}_{p,p\pi^0}} = g,\ \ \ \ \
g_{{}_{\Sigma^+,\Lambda\pi^+}} = g_{{}_{\Lambda,\Sigma^+\pi^-}} =
-\frac{2}{\sqrt 3}\alpha g,
\]

\begin{equation}
g_{{}_{\Sigma^+,\Sigma^+\pi^0}} = 2(1-\alpha)g,\ \ \ \ \
g_{{}_{\Sigma^0, pK^-}} = -g_{{}_{\Xi^0,\Xi^0\pi^0}} = (2\alpha-1)g
\label{eq2}
\end{equation}

\[ g_{{}_{\Lambda,pK^-}} = \frac{1}{\sqrt 3}(3-2\alpha)g,\ \ \ \ \
g_{{}_{\Xi^-,\Lambda K^-}} = \frac{1}{\sqrt 3}(4\alpha-3)g.
\]

\noindent
The connection between $\alpha$ and $g$ and the reduced from factors $F$ and
$D$ are $\alpha = D/(D+F)$ and $g = D+F$.

As a first approximation we shall neglect isospin violations, i.e., we
shall assume that $H_{\rm st}$ is an SU(2) scalar. However, we shall
not neglect SU(3) breaking. From Eqs.~(\ref{su2limit}) one obtains
for the $A$'s and $B$'s the results:

\[
A_1 = \delta' \sqrt 3 g^{{}^{\rm p,sp}}_{{}_{p,p\pi^0}} + \delta
(g^{{}^{\rm s,ss}}_{{}_{\Lambda,pK^-}} - g^{{}^{\rm
s,pp}}_{{}_{\Lambda,\Sigma^+\pi^-}} ) ,
\]

\[
A_2
=
-
\frac{1}{\sqrt{2}}
[
\delta'
\sqrt 3 g^{{}^{\rm p,sp}}_{{}_{p,p\pi^0}} +
\delta
(
g^{{}^{\rm s,ss}}_{{}_{\Lambda,pK^-}} -
g^{{}^{\rm s,pp}}_{{}_{\Lambda,\Sigma^+\pi^-}}
)
]
,
\]

\begin{equation}
A_3
=
\delta
(
\sqrt 2 g^{{}^{\rm s,ss}}_{{}_{\Sigma^0,p K^-}} +
\sqrt{\frac{3}{2}} g^{{}^{\rm s,pp}}_{{}_{\Sigma^+,\Lambda\pi^+}} +
\frac{1}{\sqrt{2}} g^{{}^{\rm s,pp}}_{{}_{\Sigma^+,\Sigma^+\pi^0}}
)
,
\label{cuatronl}
\end{equation}

\[
A_4
=
-\delta'
\sqrt 2 g^{{}^{\rm p,sp}}_{{}_{p,p\pi^0}} +
\delta
(
\sqrt{\frac{3}{2}} g^{{}^{\rm s,pp}}_{{}_{\Sigma^+,\Lambda\pi^+}} -
\frac{1}{\sqrt{2}} g^{{}^{\rm s,pp}}_{{}_{\Sigma^+,\Sigma^+\pi^0}}
)
,
\]

\[
A_5
=
-\delta'
g^{{}^{\rm p,sp}}_{{}_{p,p\pi^0}} -
\delta
(
g^{{}^{\rm s,ss}}_{{}_{\Sigma^0,pK^-}} +
g^{{}^{\rm s,pp}}_{{}_{\Sigma^+,\Sigma^+\pi^0}}
)
,
\]

\[
A_6
=
\delta'
g^{{}^{\rm p,sp}}_{{}_{\Sigma^+,\Lambda\pi^+}} +
\delta
(
g^{{}^{\rm s,ss}}_{{}_{\Xi^-,\Lambda K^-}} +
\sqrt 3 g^{{}^{\rm s,pp}}_{{}_{\Xi^0,\Xi^0\pi^0}}
)
,
\]

\[
A_7
=
\frac{1}{\sqrt{2}}
[
\delta'
g^{{}^{\rm p,sp}}_{{}_{\Sigma^+,\Lambda\pi^+}} +
\delta
(
g^{{}^{\rm s,ss}}_{{}_{\Xi^-,\Lambda K^-}} +
\sqrt 3 g^{{}^{\rm s,pp}}_{{}_{\Xi^0,\Xi^0\pi^0}}
)
]
,
\]

\noindent
and

\[
B_1
=
\sigma
(
- \sqrt 3 g_{{}_{p,p\pi^0}} +
g_{{}_{\Lambda,pK^-}} - g_{{}_{\Lambda,\Sigma^+\pi^-}}
)
,
\]

\[
B_2
=
-
\frac{1}{\sqrt{2}}
\sigma
(
- \sqrt 3 g_{{}_{p,p\pi^0}} +
g_{{}_{\Lambda,pK^-}} - g_{{}_{\Lambda,\Sigma^+\pi^-}}
)
,
\]

\begin{equation}
B_3
=
\sigma
(
\sqrt 2 g_{{}_{\Sigma^0,p K^-}} +
\sqrt{\frac{3}{2}} g_{{}_{\Sigma^+,\Lambda\pi^+}} +
\frac{1}{\sqrt{2}} g_{{}_{\Sigma^+,\Sigma^+\pi^0}}
)
,
\label{cinconl}
\end{equation}

\[
B_4
=
\sigma
(
\sqrt 2 g_{{}_{p,p\pi^0}} +
\sqrt{\frac{3}{2}} g_{{}_{\Sigma^+,\Lambda\pi^+}} -
\frac{1}{\sqrt{2}} g_{{}_{\Sigma^+,\Sigma^+\pi^0}}
)
,
\]

\[
B_5
=
\sigma
(
g_{{}_{p,p\pi^0}} -
g_{{}_{\Sigma^0,pK^-}} - g_{{}_{\Sigma^+,\Sigma^+\pi^0}}
)
,
\]

\[
B_6
=
\sigma
(
- g_{{}_{\Sigma^+,\Lambda\pi^+}} +
g_{{}_{\Xi^-,\Lambda K^-}} +
\sqrt 3 g_{{}_{\Xi^0,\Xi^0\pi^0}}
)
,
\]

\[
B_7
=
\frac{1}{\sqrt{2}}
\sigma
(
- g_{{}_{\Sigma^+,\Lambda\pi^+}} +
g_{{}_{\Xi^-,\Lambda K^-}} +
\sqrt 3 g_{{}_{\Xi^0,\Xi^0\pi^0}}
)
.
\]

From the above results one readily obtains the equalities:

\begin{equation}
A_2 = -\frac{1}{\sqrt 2} A_1,\ \ \ \ \ \
A_5 = \frac{1}{\sqrt 2} (A_4-A_3),\ \ \ \ \ \
A_7 = \frac{1}{\sqrt 2} A_6,
\label{seisA}
\end{equation}

\begin{equation}
B_2 = -\frac{1}{\sqrt 2} B_1,\ \ \ \ \ \
B_5 = \frac{1}{\sqrt 2} (B_4-B_3),\ \ \ \ \ \
B_7 = \frac{1}{\sqrt 2} B_6.
\label{siete}
\end{equation}

\noindent
These are the predictions of the $|\Delta I|=1/2$
rule~\cite{gellmann55,marshak69}. That is, mirror matter mixings in
hadrons as introduced above lead to the predictions of the $|\Delta
I|=1/2$ rule, but notice that they do not lead to the $|\Delta I|=1/2$
rule itself. This rule originally refers to the isospin covariance
properties of the effective non-leptonic interaction Hamiltonian to be
sandwiched between strong-flavor and parity eigenstates. The $I=1/2$ part
of this Hamiltonian is en\-han\-ced over the $I=3/2$ part. In contrast, in
the case of mirror matter admixtures $H_{\rm st}$ has been assumed
to be isospin invariant, i.e., in this case the rule should be called a
$\Delta I=0$ rule.

It must be stressed that the results~(\ref{seisA}) and
(\ref{siete}) are very general: (i) the predictions of the
$|\Delta I|=1/2$ rule are obtained simultaneously for the $A$
and $B$ amplitudes, (ii) they are independent of the mixing
angles $\sigma$, $\delta$, and $\delta'$, and (iii) they are
also independent of particular values of the YCC. They will be
violated by isospin breaking corrections. So, they should be
quite accurate, as is experimentally the case.

Now, we assume that $H_{\rm st}$ is an invariant operator, an SU(3)
invariant in the case of Eqs.~(\ref{cuatronl}) and (\ref{cinconl}).
We replace the symmetry limit values of the $g$'s,
Eqs.~(\ref{eq2}), in (\ref{cuatronl}) and (\ref{cinconl}).
In this case one obtains

\[
A_1 = -\sqrt{2}A_2 = \frac{1}{\sqrt 3} \left\{ \delta' 3
g^{{}^{\rm p,sp}} + \delta \left[ (3-2\alpha)g^{{}^{\rm s,ss}} +
2\alpha g^{{}^{\rm s,pp}} \right] \right\},
\]

\begin{equation}
A_3
=
\delta \sqrt{2} (2\alpha - 1)
(g^{{}^{\rm s,ss}} - g^{{}^{\rm s,pp}})
,
\label{asu3nl}
\end{equation}

\[
A_4
=
-\sqrt{2} (\delta' g^{{}^{\rm p,sp}} + \delta g^{{}^{\rm s,pp}})
,\qquad A_5 = \frac{1}{\sqrt{2}} (A_4 - A_3),
\]

\[
A_6 = \sqrt{2} A_7 = \frac{1}{\sqrt{3}}
\left\{ -\delta' 2 \alpha g^{{}^{\rm p,sp}} +
\delta \left[(4\alpha - 3) g^{{}^{\rm s,ss}} -
3 (2\alpha - 1) g^{{}^{\rm s,pp}}\right]\right\},
\]

\noindent
and

\begin{equation}
B_1 = B_2 = \cdots = B_7 = 0.
\label{bsu3nl}
\end{equation}

\noindent
The so-called parity-conserving amplitudes automatically vanish
in the SU(3) symmetry limit~\cite{symmetry}.

For the so-called parity-violating amplitudes to vanish in the
same limit, one observes in Eq.~(\ref{asu3nl}) that two
conditions must be met, namely

\begin{equation}
-g^{{}^{\rm p,sp}}=g^{{}^{\rm s,pp}}=g^{{}^{\rm s,ss}}\quad
{\rm and}\quad\delta'=\delta.
\label{cond}
\end{equation}

\noindent
That is, when these two conditions are satisfied one has

\begin{equation}
A_1 = A_2 = \cdots = A_7 = 0.
\label{asu3limitnl}
\end{equation}

\section{Two-body non-leptonic decay amplitudes of
$\bm{\Omega^-}$} \label{onl}

We shall study now the decays $\Omega^-\to\Xi^-\pi^0$,
$\Omega^-\to\Xi^0\pi^-$, and $\Omega^-\to\Lambda K^-$, not
considered before~\cite{symmetry}. They are described by a
Lorentz invariant amplitude of the form

\begin{equation}
\langle M(q) B'(p') |H_{\rm st}| B(p) \rangle =
\bar{u}(p') ( {\cal B} + \gamma^5 {\cal C} ) q^{\mu} u_{\mu}(p),
\label{amplitude}
\end{equation}

\noindent
where $u$ and $u_{\mu}$ are Dirac and Rarita-Schwinger spinors,
respectively, and ${\cal B}$ and ${\cal C}$ are $p$-wave
(parity-conserving) and $d$-wave (parity-violating) amplitudes,
respectively. $H_{\rm st}$ is the transition operator and
$B(p)$, $B'(p')$, and $M(q)$ represent $s=3/2$, $s=1/2$ baryons,
and a pseudoscalar meson, respectively.

In our approach the transition operator $H_{\rm st}$ is the strong
flavor and parity conserving part of the Hamiltonian responsible for
the two-body strong decays of the other $s=3/2$ resonances in the
decuplet where $\Omega^-_{\rm s}$ belongs to.

To obtain explicit expressions for ${\cal B}$ and ${\cal C}$ we
need the mirror admixtures in $\Omega^-_{\rm ph}$, $\pi^0_{\rm
ph}$, $\pi^-_{\rm ph}$, $K^-_{\rm ph}$, $\Lambda_{\rm ph}$,
$\Xi^0_{\rm ph}$, and $\Xi^-_{\rm ph}$. We take them from our
previous work~\cite{omegasymlim},

\begin{equation}
\Omega^-_{\rm ph} =
\Omega^-_{\rm s} -
\sigma\sqrt{3}\Xi^{*-}_{\rm s} +
\delta'\sqrt{3}\Xi^{*-}_{\rm p} +
\cdots,
\label{omega}
\end{equation}

\noindent
and Eqs.~(\ref{mph}) and (\ref{bph}).

The properties of $H_{\rm st}$ discussed above, lead to

\[
{\cal B}(\Omega^-\to\Xi^-\pi^0) =
-\sigma
(\sqrt 3 g^{{}^{\rm s,sp}}_{{}_{\Xi^{*-},\Xi^-\pi^0}} +
\frac{1}{\sqrt{2}} g^{{}^{\rm
s,sp}}_{{}_{\Omega^-,\Xi^-\bar{K}^0}}),
\]

\begin{equation}
{\cal B}(\Omega^-\to\Xi^0\pi^-) =
\sigma
(
- \sqrt 3 g^{{}^{\rm s,sp}}_{{}_{\Xi^{*-},\Xi^0\pi^-}} +
g^{{}^{\rm s,sp}}_{{}_{\Omega^-,\Xi^0K^-}}
) ,
\label{bxi0}
\end{equation}

\[
{\cal B}(\Omega^-\to\Lambda K^-) =
\sigma
(
- \sqrt 3 g^{{}^{\rm s,sp}}_{{}_{\Xi^{*-},\Lambda K^-}} +
\sqrt{\frac{3}{2}}g^{{}^{\rm s,sp}}_{{}_{\Omega^-,\Xi^0K^-}}
),
\]

\[
{\cal C}(\Omega^-\to\Xi^-\pi^0) =
\delta'
\sqrt{3} g^{{}^{\rm p,sp}}_{{}_{\Xi^{*-},\Xi^-\pi^0}} -
\delta
\frac{1}{\sqrt{2}}g^{{}^{\rm s,ss}}_{{}_{\Omega^-,\Xi^-\bar{K}^0}},
\]

\begin{equation}
{\cal C}(\Omega^-\to\Xi^0\pi^-) =
\delta'
\sqrt{3} g^{{}^{\rm p,sp}}_{{}_{\Xi^{*-},\Xi^0\pi^-}} +
\delta
g^{{}^{\rm s,ss}}_{{}_{\Omega^-,\Xi^0 K^-}},
\label{cxi0}
\end{equation}

\[
{\cal C}(\Omega^-\to\Lambda K^-) =
\delta'
\sqrt{3} g^{{}^{\rm p,sp}}_{{}_{\Xi^{*-},\Lambda K^-}} +
\delta
\sqrt{\frac{3}{2}}g^{{}^{\rm s,pp}}_{{}_{\Omega^-,\Xi^0K^-}}.
\]

\noindent
The constants $g^{{}^{\rm s,sp}}_{{}_{B,B'M}}$ are the Yukawa strong
couplings observed in the strong two-body decays of $s=3/2$ resonances.
The constants $g^{{}^{\rm p,sp}}_{{}_{B,B'M}}$, $g^{{}^{\rm
s,ss}}_{{}_{B,B'M}}$ and $g^{{}^{\rm s,pp}}_{{}_{B,B´M}}$ are new, because
they involve mirror matter. As for NLDH, they are defined by the matrix
elements of $H_{\rm st}$ between flavor and parity eigenstates. In the
isospin limit they are related by

\[
g_{{}_{\Omega^-,\Xi^0 K^-}}=g_{{}_{\Omega^-,\Xi^-\bar{K}^0}},
\]

\begin{equation}
g_{{}_{\Xi^{*-},\Xi^0\pi^-}}=-\sqrt{2}g_{{}_{\Xi^{*-},\Xi^-\pi^0}},
\label{su2b}
\end{equation}

\noindent
In the SU(3) limit one also has

\[
g_{{}_{\Xi^{*-},\Xi^-\pi^0}} = -\frac{1}{\sqrt{6}}
g_{{}_{\Omega^-,\Xi^-\bar{K}^0}},
\]

\begin{equation}
g_{{}_{\Xi^{*-},\Lambda K^-}} =
\frac{1}{\sqrt{2}} g_{{}_{\Omega^-,\Xi^-\bar{K}^0}},
\label{su3b}
\end{equation}

\noindent
where the indices s and p may be dropped out.

In the explicit amplitudes (\ref{bxi0}) and (\ref{cxi0}) we can now replace
the isospin equalities (\ref{su2b}). Eqs.~(\ref{bxi0}) and (\ref{cxi0})
become

\[
{\cal B}(\Omega^-\to\Xi^-\pi^0) =
-\sigma
(
 \sqrt 3 g_{{}_{\Xi^{*-},\Xi^-\pi^0}} +
\frac{1}{\sqrt{2}}g_{{}_{\Omega^-,\Xi^-\bar{K}^0}}
),
\]

\begin{equation}
{\cal B}(\Omega^-\to\Xi^0\pi^-) =
\sigma
(
\sqrt{6}  g_{{}_{\Xi^{*-},\Xi^-\pi^0}} +
g_{{}_{\Xi^-,\Omega^-\bar{K}^0}}
),
\label{bxi02}
\end{equation}

\[
{\cal B}(\Omega^-\to\Lambda K^-) =
\sigma
(
- \sqrt 3 g_{{}_{\Xi^{*-},\Lambda K^-}} +
\sqrt{\frac{3}{2}}g_{{}_{\Omega^-,\Xi^-\bar{K}^0}}
),
\]

\[
{\cal C}(\Omega^-\to\Xi^-\pi^0) =
\delta'
\sqrt{3} g^{{}^{\rm p,sp}}_{{}_{\Xi^{*-},\Xi^-\pi^0}} -
\delta
\frac{1}{\sqrt{2}}g^{{}^{\rm s,ss}}_{{}_{\Omega^-,\Xi^-\bar{K}^0}},
\]

\begin{equation}
{\cal C}(\Omega^-\to\Xi^0\pi^-) =
- \delta'
\sqrt{6} g^{{}^{\rm p,sp}}_{{}_{\Xi^{*-},\Xi^-\pi^0}} +
\delta
g^{{}^{\rm s,ss}}_{{}_{\Omega^-,\Xi^-\bar{K}^0}},
\label{cxi02}
\end{equation}

\[
{\cal C}(\Omega^-\to\Lambda K^-) =
\delta'
\sqrt{3} g^{{}^{\rm p,sp}}_{{}_{\Xi^{*-},\Lambda K^-}} +
\delta
\sqrt{\frac{3}{2}}g^{{}^{\rm s,pp}}_{{}_{\Omega^-,\Xi^-\bar{K}^0}}.
\]

\noindent
We have omitted the indices s and p in the $g$'s of the ${\cal B}$
amplitudes because the states involved carry the normal intrinsic parities
of hadrons.

One readily sees that the equalities

\begin{equation}
\frac{{\cal B}(\Omega^-\to\Xi^-\pi^0) }{{\cal B}(\Omega^-\to\Xi^0\pi^-)} =
\frac{{\cal C}(\Omega^-\to\Xi^-\pi^0) }{{\cal C}(\Omega^-\to\Xi^0\pi^-)} =
-\frac{1}{\sqrt{2}}
\label{deltai}
\end{equation}

\noindent
are obtained. This is the $|\Delta I|=1/2$ rule in these decay
modes and applies to both decay amplitudes. The mirror admixtures in
$\Omega^-$ lead to the results of this rule. However, it must be recalled
that in this approach these results are really a $\Delta I=0$ rule, since
isospin is assumed unbroken.

If we now assume the SU(3) limit, we can replace (\ref{su3b}) in the
amplitudes (\ref{bxi02}) and (\ref{cxi02}). In this case one obtains

\begin{equation}
{\cal B}(\Omega^-\to\Xi^-\pi^0) =
{\cal B}(\Omega^-\to\Xi^0\pi^-) = {\cal B}(\Omega^-\to\Lambda K^-) = 0,
\label{su3lmt1}
\end{equation}

\[
{\cal C}(\Omega^-\to\Xi^-\pi^0) =
- \frac{1}{\sqrt{2}}{\cal C}(\Omega^-\to\Xi^0\pi^-) =
- \frac{1}{\sqrt{2}}
(\delta'
g^{{}^{\rm p,sp}}_{{}_{\Omega^-,\Xi^-\bar{K}^0}} +
\delta
g^{{}^{\rm s,ss}}_{{}_{\Omega^-,\Xi^-\bar{K}^0}}),
\]

\begin{equation}
{\cal C}(\Omega^-\to\Lambda K^-) =
\sqrt{\frac{3}{2}}
(\delta'
g^{{}^{\rm p,sp}}_{{}_{\Omega^-,\Xi^-\bar{K}^0}} +
\delta
g^{{}^{\rm s,pp}}_{{}_{\Omega^-,\Xi^-\bar{K}^0}}).
\label{su3lmt3}
\end{equation}

\noindent
We observe from these equations that, as was the case in
NLDH, the so-called parity-conserving
amplitudes vanish automatically in this limit, as previously
reported~\cite{omegasymlim}. Now, one can also observe for the
so-called parity-violating amplitudes that

\begin{equation}
{\cal C}(\Omega^-\to\Xi^-\pi^0) =
{\cal C}(\Omega^-\to\Xi^0\pi^-) =
{\cal C}(\Omega^-\to\Lambda K^-)=0
\label{csu3lmt}
\end{equation}

\noindent
when the same conditions of Eq.~(\ref{cond}) are met.

\section{Weak radiative decay amplitudes of hyperons}
\label{wrdh}

Next, we shall discuss the application of the above scheme to the
weak radiative decays of hyperons~\cite{universality}. In this
case, these transitions receive contributions due to the matrix
elements of the parity and flavor conserving electromagnetic
part $H_{\rm em} = iJ^\mu_{\rm em}A_\mu$ of the Hamiltonian
between the physical baryons with mirror matter admixtures,
Eqs.~(\ref{bph}). $A_\mu$ represents the photon field operator
and $J^\mu_{\rm em}$ is the ordinary electromagnetic current
operator. This operator is --- and this must be emphasized --- a
flavor-conserving Lorentz proper four-vector. The radiative decay
amplitudes we want are given by the usual matrix elements
$\langle\gamma,B'_{\rm ph}|H_{\rm em}|B_{\rm ph}\rangle$, where
$B_{\rm ph}$ and $B'_{\rm ph}$ stand for hyperons. The hadronic
parts of the transition amplitudes are then given by:

\[
\langle p_{\rm ph} | J^{\mu}_{\rm em} | \Sigma^+_{\rm ph} \rangle =
\bar u_p [
\sigma (f^p_{\rm 2\,ss} - f^{\Sigma^+}_{\rm 2\,ss}) -
(\delta' f^p_{\rm 2\,sp} +
\delta f^{\Sigma^+}_{\rm 2\,ps})
\gamma^5] i\sigma^{\mu\nu}q_{\nu} u_{\Sigma^+}
\]

\[
\langle \Sigma^-_{\rm ph} | J^{\mu}_{\rm em} | \Xi^-_{\rm ph} \rangle =
\bar u_{\Sigma^-} [
\sigma (-f^{\Sigma^-}_{\rm 2\,ss} + f^{\Xi^-}_{\rm 2\,ss} ) +
(\delta' f^{\Sigma^-}_{\rm 2\,sp} +
\delta f^{\Xi^-}_{\rm 2\,ps})
\gamma^5]i\sigma^{\mu\nu}q_{\nu} u_{\Xi^-}
\]

\begin{eqnarray}
\langle n_{\rm ph} | J^{\mu}_{\rm em} | \Lambda_{\rm ph} \rangle
& = &
\bar u_n \left\{
\sigma \left[
\sqrt{\frac{3}{2}} (-f^n_{\rm 2\,ss} + f^{\Lambda}_{\rm 2\,ss}) +
\frac{1}{\sqrt 2} f^{\Sigma^0\Lambda}_{\rm 2\,ss}
\right]\right.
\nonumber \\
& &
\left. + \left[ \sqrt{\frac{3}{2}}
(\delta' f^n_{\rm 2\,sp}  +
\delta f^{\Lambda}_{\rm 2\,ps}) +
\delta \frac{1}{\sqrt 2} f^{\Sigma^0\Lambda}_{\rm 2\,ps}
\right]\gamma^5\right\}i\sigma^{\mu\nu}q_{\nu} u_{\Lambda}
\label{seis}
\end{eqnarray}

\begin{eqnarray}
\langle \Lambda_{\rm ph} | J^{\mu}_{\rm em} | \Xi^0_{\rm ph} \rangle
& = &
\bar u_{\Lambda} \left\{
\sigma \left[ \sqrt{\frac{3}{2}}
(-f^{\Lambda}_{\rm 2\,ss} + f^{\Xi^0}_{\rm 2\,ss}) -
\frac{1}{\sqrt 2} f^{\Sigma^0\Lambda}_{\rm 2\,ss}
\right]\right.
\nonumber \\
& &
\left. + \left[ \sqrt{\frac{3}{2}}
(\delta' f^{\Lambda}_{\rm 2\,sp} +
\delta f^{\Xi^0}_{\rm 2\,ps}) +
\delta' \frac{1}{\sqrt 2} f^{\Sigma^0\Lambda}_{\rm 2\,sp}
\right]\gamma^5\right\}i\sigma^{\mu\nu}q_{\nu} u_{\Xi^0}
\nonumber
\end{eqnarray}

\begin{eqnarray}
\langle \Sigma^0_{\rm ph} | J^{\mu}_{\rm em} | \Xi^0_{\rm ph} \rangle
& = &
\bar u_{\Sigma^0} \left\{
\sigma \left[
\frac{1}{\sqrt 2} (-f^{\Sigma^0}_{\rm 2\,ss} + f^{\Xi^0}_{\rm 2\,ss}) -
\sqrt{\frac{3}{2}} f^{\Sigma^0\Lambda}_{\rm 2\,ss}
\right]\right.
\nonumber \\
& &
\left. + \left[ \frac{1}{\sqrt 2} (
\delta' f^{\Sigma^0}_{\rm 2\,sp} +
\delta f^{\Xi^0}_{\rm 2\,ps}) +
\delta' \sqrt{\frac{3}{2}}f^{\Sigma^0\Lambda}_{\rm 2\,sp}
\right]\gamma^5\right\}i\sigma^{\mu\nu}q_{\nu} u_{\Xi^0}
\nonumber
\end{eqnarray}

\noindent
In these amplitudes only contributions to first order in $\sigma$,
$\delta$, and $\delta'$ need be kept. Each matrix element of $J^\mu_{\rm
em}$ between flavour and parity eigenstates is flavour and parity
conserving and can be expanded in terms of charge $f_1(0)$ form factors
and anomalous magnetic $f_2(0)$ form factors. We have used the generator
properties of the electric charge, which require $f^p_{\rm
1\,ss}=f^{\Sigma^+}_{\rm 1\,ss}=1$, etc. and also, since s and p baryons
belong to different irreducible representations, $f^p_{\rm
1\,sp}=f^{\Sigma^+}_{\rm 1\,sp}=0$, etc. Notice that the
amplitudes~(\ref{seis}) are all of the form
$\bar{u}_{B'}(C+D\gamma_5)i\sigma^{\mu\nu}q_{\nu}u_B$, where the spinors
$u_B$, $B=p,\Sigma^+$, etc. are ordinary four-component Dirac spinors and
$q=p_{B'}-p_B$. $C$ is the so-called parity-conserving amplitude
and $D$ is the so-called parity-violating one. We stress,
however, that in this model both $C$ and $D$ are parity and
flavor conserving.

In the SU(3) symmetry limit the anomalous magnetic moments $f_2$
are related by

\[
f^{\Sigma^+}_2 = f_2^p,\ \ \ \
f^{\Xi^-}_2 = f^{\Sigma^-}_2,\ \ \ \
f^{\Xi^0}_2 = f^n_2,
\]

\begin{equation}
f^{\Sigma^0\Lambda}_2 = \frac{\sqrt 3}{2}f^n_2,\ \ \ \
f^{\Sigma^0}_2 =  -\frac{1}{2}f^n_2,\ \ \ \
f^\Lambda_2 = \frac{1}{2}f^n_2.
\label{eq4}
\end{equation}

\noindent
When Eqs.~(\ref{eq4}) are replaced into Eqs.~(\ref{seis}) we obtain

\begin{equation}
C(\Sigma^+\to p\gamma)=C(\Xi^-\to \Sigma^-\gamma)=
C(\Lambda\to n\gamma)=C(\Xi^0\to \Lambda\gamma)=
C(\Xi^0\to \Sigma^-\gamma)=0,
\label{Csu3}
\end{equation}

\[
D(\Sigma^+\to p\gamma)=
-\delta' f^p_{\rm 2\,sp} -
\delta f^p_{\rm 2\,ps},
\qquad
D(\Xi^-\to \Sigma^-\gamma)=
\delta' f^{\Sigma^-}_{\rm 2\,sp} +
\delta f^{\Sigma^-}_{\rm 2\,ps},
\]

\begin{equation}
D(\Lambda\to n\gamma)=D(\Xi^0\to \Lambda\gamma)=
\sqrt{3}D(\Xi^0\to \Sigma^0\gamma)=
\sqrt{\frac{3}{2}}
(\delta' f^n_{\rm 2\,sp} +
\delta f^n_{\rm 2\,ps}),
\label{Dsu3}
\end{equation}

\noindent
From Eqs.~(\ref{Csu3}) one sees that the same result of NLDH and
$\Omega^-$ two-body non-leptonic decays for the so-called
parity-conserving amplitudes is automatically
obtained~\cite{symmetry}.

The so-called parity-violating amplitudes Eqs.~(\ref{Dsu3})
vanish in the SU(3) symmetry limit,

\begin{equation}
D(\Sigma^+\to p\gamma)=D(\Xi^-\to \Sigma^-\gamma)=
D(\Lambda\to n\gamma)=D(\Xi^0\to \Lambda\gamma)=0,
\label{Dlimsu3}
\end{equation}

\noindent
when the conditions

\begin{equation}
\delta'=\delta\quad
{\rm and}\quad f^B_{\rm 2\,ps}=-f^B_{\rm 2\,sp}
\label{condwrd}
\end{equation}

\noindent
are fulfilled. Notice that Eq.~(\ref{condwrd}) is the
equivalent of Eq.~(\ref{cond}), except that now it is the $H_{\rm
em}$ part of the Hamiltonian that is involved.

\section{Two photon decay amplitudes of $\bm{K_{\rm L}}$ and
$\bm{K_{\rm S}}$} \label{ksl}

Let us now study the symmetry limit properties of the decays
$K_{\rm L,S}\to\gamma\gamma$~\cite{kl2g,ks2g}. Our
phenomenological model, based on parity and flavor admixtures of
mirror matter in ordinary mesons of Eqs.~(\ref{mph}), may
contribute to the enhancement phenomenon observed in these
decays via the electromagnetic part $H_{\rm em}$ of the
Hamiltonian transition operator.

The physical mesons (mass eigenstates) of Eqs.~(\ref{mph})
are used to form the short and long lived $K$'s, $K_{\rm L,S}$.
Notice that the physical mesons satisfy ${\rm CP}K^0_{\rm
ph}=-\bar{K}^0_{\rm ph}$ and ${\rm CP}\bar{K}^0_{\rm
ph}=-K^0_{\rm ph}$. We can form the CP-eigenstates $K_{\rm 1}$
and $K_{\rm 2}$ as

\begin{equation}
K_{\rm 1_{ph}} = \frac{1}{\sqrt{2}} (K^0_{\rm ph} - \bar{K}^0_{\rm ph})
\qquad\mbox{and}\qquad
K_{\rm 2_{ph}} = \frac{1}{\sqrt{2}} (K^0_{\rm ph} + \bar{K}^0_{\rm ph}),
\label{dos}
\end{equation}

\noindent
the $K_{\rm 1_{ph}}$ ($K_{\rm 2_{ph}}$) is an even (odd) state with respect
to CP. Here, we shall not consider CP-violation and therefore,
$|K_{\rm S,L}\rangle = |K_{1,2}\rangle$.

Substituting the expressions given in Eqs.~(\ref{mph}), we obtain,

\[
K_{\rm L_{ph}} =
K_{\rm L_p} +
\sigma \pi^0_{\rm p} +
\sqrt{3} \sigma \eta_{\rm 8\,p} ,
\]

\begin{equation}
K_{\rm S_{ph}} =
K_{\rm S_p} +
\frac{1}{\sqrt{3}} (2\delta + \delta') \eta_{\rm 8\,s} +
\delta' \pi^0_{\rm s} +
\sqrt{\frac{2}{3}} (\delta - \delta') \eta_{\rm 1\,s} ,
\label{tres}
\end{equation}

\noindent
where the usual definitions
$K_{\rm 1_p} = (K^0_{\rm p} - \bar{K}^0_{\rm p})/\sqrt{2}$
and
$K_{\rm 2_p} = (K^0_{\rm p} + \bar{K}^0_{\rm p})/\sqrt{2}$
were used.

A very simple calculation, using Eqs.~(\ref{tres}) and $H_{\rm
em}$, gives the following contributions of mirror matter
admixtures to the $K_{\rm L,S}\to\gamma\gamma$ amplitudes:

\begin{equation}
F_{K_{\rm L}\gamma\gamma} =
\sigma F_{\pi^0_{\rm p}\gamma\gamma} +
\sqrt{3}\sigma F_{\eta_{\rm 8\,p}\gamma\gamma},
\label{cinco}
\end{equation}

\begin{equation}
F_{K_{\rm S}\gamma\gamma} =
\frac{1}{\sqrt{3}}(2\delta+\delta')F_{\eta_{\rm 8\,s}\gamma\gamma} +
\delta' F_{\pi^0_{\rm s}\gamma\gamma} +
\sqrt{\frac{2}{3}}(\delta-\delta')F_{\eta_{\rm 1\,s}\gamma\gamma},
\label{cuatro}
\end{equation}

\noindent
where
$F_{K_{\rm L}\gamma\gamma}=
\langle\gamma\gamma|H_{\rm em}|K_{\rm L_{ph}}\rangle$,
$F_{K_{\rm S}\gamma\gamma}=
\langle\gamma\gamma|H_{\rm em}|K_{\rm S_{ph}}\rangle$,
and
$F_{\pi^0_{\rm p}\gamma\gamma}=\langle\gamma\gamma
|H_{\rm em}|\pi^0_{\rm p}\rangle$, etc.

Given that $K_{\rm S}$ and $K_{\rm L}$ are ${\rm CP}=+1$ and ${\rm CP}=-1$
pure states respectively, and because the two-photon state is a ${\rm C}=+1$
state, then $K_{\rm S}\to\gamma\gamma$ must go through a so-called
parity-violating transition while $K_{\rm L}\to\gamma\gamma$ goes through
a parity-conserving transition. In the first case the two-photon final
state is ${\rm P}=+1$ while in the second one, ${\rm P}=-1$. However, as
we can see from Eqs.~(\ref{cinco}) and (\ref{cuatro}), in the context of
mirror matter admixtures all the contributions to both amplitudes are
flavour and parity conserving.

In the SU(3) symmetry limit (U-spin invariance) one has

\begin{equation}
F_{\eta_8\gamma\gamma}=-\frac{1}{\sqrt{3}}F_{\pi^0\gamma\gamma}.
\label{cinco1}
\end{equation}

\noindent
The amplitudes (\ref{cinco}) and (\ref{cuatro}) become in this
limit

\begin{equation}
F_{K_{\rm L}\gamma\gamma} = 0
\label{fklsu3}
\end{equation}

\noindent
and

\begin{equation}
F_{K_{\rm S}\gamma\gamma} = \sqrt{\frac{2}{3}} (\delta'-\delta)
(\sqrt{\frac{2}{3}}F_{\pi^0_{\rm s}\gamma\gamma} - F_{\eta_{\rm
1s}\gamma\gamma}).
\label{fkssu3}
\end{equation}

\noindent
Once again, the so-called parity-conserving amplitude of $K_{\rm
L}\to\gamma\gamma$ vanishes automatically.

The vanishing of the so-called parity-violating amplitude

\begin{equation}
F_{K_{\rm S}\gamma\gamma} = 0,
\label{fkssu3limit}
\end{equation}

\noindent
requires only that $\delta'=\delta$ of Eqs.~(\ref{cond}) and
(\ref{condwrd}) be satisfied. The reason for this simplification
is that only one hadron is involved in these $K_{\rm
L,S}\to\gamma\gamma$ decays.

\section{$\bm{K\to\pi\pi}$ decay amplitudes}
\label{ktopipi}

Contributions of the parity and flavor admixtures in physical
mesons to the $K\to\pi\pi$ decay amplitudes are
determined by the matrix elements of the flavor and parity
conserving strong part of the Hamiltonian, $H_{\rm
st}$\ ~\cite{kpipi}.

From expressions~(\ref{tres}) of Sec.~\ref{ksl} we obtain
some important conclusions. Since the Hamiltonian $H_{\rm st}$ is
by assumption isoscalar and also a flavour and parity conserving
one, we notice that the physical state $K_{\rm S_{ph}}$ can only
decay into two pions and not into three pions. In the latter
case, the final state made out of three pions has total angular
momentum equal to zero. The parity is odd, since each of the
pions has negative parity, and then the Hamiltonian can not make
the transition. In the case of having two pions in the final
state, the transition is possible and proportional to the
constants $\delta$ and $\delta'$. Similarly for the state
$K_{\rm L_{ph}}$: it has to go to three pions and the amplitude
is proportional to $\sigma$. The above qualitative behavior is
observed experimentally neglecting CP-violation effects.
 
The amplitudes for the decays $K\rightarrow\pi\pi$ are denoted by
$A_{+0} = \langle\pi^+_{\rm ph}\pi^0_{\rm ph} |H_{\rm st}|
K^+_{\rm ph}\rangle$, $A_{+-} = \langle\pi^+_{\rm ph}\pi^-_{\rm
ph} |H_{\rm st}| K_{\rm S_{ph}}\rangle$, and $A_{00} =
\langle\pi^0_{\rm ph}\pi^0_{\rm ph} |H_{\rm st}| K_{\rm
S_{ph}}\rangle$. After the substitution of the physical mass
eigenstates given in Eqs.~(\ref{mph}), we obtain explicitly

\[
A_{+0} =
-\delta' \langle\pi^+_{\rm p}\pi^0_{\rm p} |H_{\rm st}|
\pi^+_{\rm s}\rangle + \frac{1}{\sqrt{2}}\delta
\langle\pi^+_{\rm p} K^0_{\rm s}|H_{\rm st}|K^+_{\rm
p}\rangle -\delta \langle K^+_{\rm s}\pi^0_{\rm p} |H_{\rm
st}| K^+_{\rm p}\rangle
\]

\begin{eqnarray}
\label{cuatrom}
A_{+-} &=&
\frac{1}{\sqrt{3}}(2\delta+\delta')
\langle\pi^+_{\rm p}\pi^-_{\rm p} |H_{\rm st}| \eta_{\rm 8s}\rangle
+\delta' \langle\pi^+_{\rm p}\pi^-_{\rm p}
|H_{\rm st}| \pi^0_{\rm s}\rangle
+\sqrt{\frac{2}{3}}(\delta-\delta')
\langle\pi^+_{\rm p}\pi^-_{\rm p} |H_{\rm st}| {\eta_{\rm
1s}}\rangle
\nonumber\\ && -\frac{1}{\sqrt{2}}\delta \langle
K^+_{\rm s}\pi^-_{\rm p} |H_{\rm st}| K^0_{\rm p}\rangle
-\frac{1}{\sqrt{2}}\delta \langle\pi^+_{\rm p}
K^-_{\rm s} |H_{\rm st}| \bar{K}^0_{\rm p}\rangle
\end{eqnarray}

\begin{eqnarray*}
A_{00} &=& \frac{1}{\sqrt{3}}(2\delta+\delta') \langle\pi^0_{\rm
p}\pi^0_{\rm p} |H_{\rm st}| \eta_{\rm 8s}\rangle
+\delta' \langle\pi^0_{\rm p}\pi^0_{\rm p} |H_{\rm
st}| \pi^0_{\rm s}\rangle
+\sqrt{\frac{2}{3}}(\delta-\delta') \langle\pi^0_{\rm
p}\pi^0_{\rm p} |H_{\rm st}| {\eta_{\rm
1s}}\rangle \\ &&
+\frac{1}{2}\delta \langle\pi^0_{\rm p} K^0_{\rm s} |H_{\rm
st}| K^0_{\rm p}\rangle +\frac{1}{2}\delta
\langle\pi^0_{\rm p}\bar{K}^0_{\rm s} |H_{\rm
st}| \bar{K}^0_{\rm p}\rangle +\frac{1}{2}\delta \langle
K^0_{\rm s}\pi^0_{\rm p} |H_{\rm st}| K^0_{\rm p}\rangle
+\frac{1}{2}\delta \langle\bar{K}^0_{\rm s}\pi^0_{\rm p}
|H_{\rm st}| \bar{K}^0_{\rm p}\rangle
\end{eqnarray*}

\noindent
In the right hand side of these equations, the amplitudes are
flavor and parity conserving. The two pions in the physical
final state of these amplitudes are either in the $I=0$ or in
the $I=2$ isospin configuration, since the $I=1$ state is
forbidden by the generalized Bose principle~\cite{neubert}.
Also, there is no contribution from the $I=2$ component since
the interaction Hamiltonian is an isosinglet (so, amplitudes
with a $\pi$ ($I=1$) in the initial state vanish). Therefore, we
can write such amplitudes in terms of a single strong coupling
constant and take into account the final state interaction by
introducing a multiplicative phase factor. We will denote each
amplitude in the form, $\langle M^{1}_{\rm i}
M^{2}_{\rm j}|H_{\rm st}|M^{3}_{\rm k}\rangle=
G^{\rm k,ij}_{M^3,M^1M^2}e^{i\alpha}$.

The amplitudes for the three decays considered become

\[
A_{+0}=
- \delta(\frac{1}{\sqrt{2}}G^{\rm p,ps}_{K^+,\pi^+ K^0}
- G^{\rm p,sp}_{K^+,K^+\pi^0})e^{i\alpha_{0}}
\]

\begin{equation}
A_{+-}=
[\frac{1}{\sqrt{3}}(2\delta + \delta') G^{\rm
s,pp}_{\eta_8,\pi^+\pi^-} +
\sqrt{\frac{2}{3}}(\delta-\delta')G^{\rm
s,pp}_{\eta_1,\pi^+\pi^-}]e^{i\alpha_{1}} \label{five}
\end{equation}

\[
A_{00}=
[\frac{1}{\sqrt{3}}(2\delta+\delta')G^{\rm
s,pp}_{\eta_8,\pi^0\pi^0} +
\sqrt{\frac{2}{3}}(\delta-\delta')G^{\rm
s,pp}_{\eta_1,\pi^0\pi^0}]e^{i\alpha_{1}} \]

\noindent
Above we have used the assumption that the strong coupling
constants have the property, $(G^{\rm
i,jk}_{M^3,M^1M^2})^{\rm CPT}=
G^{\rm i,jk}_{\overline{M}^3,\overline{M}^1 \overline{M}^2}$. Also,
we have used the properties, $\langle
j_1j_2m_1m_2|j_1j_2JM\rangle = (-1)^{J-j_1-j_2}\langle
j_2j_1m_2m_1|j_2j_1JM\rangle$ of the SU(2) Clebsch-Gordan
coefficients to simplify the expressions for the amplitudes. The
phase introduced for the final state interaction depends only on
the total isospin of the final particles; it is for this reason
that $A_{+-}$ and $A_{00}$ have the same phase factor.

From the Clebsch-Gordan Tables we get that in the SU(2)
symmetry limit $G_{K^+,\pi^+K^0} = \sqrt{2}\,G_{K^+,K^+\pi^0}$,
$G_{\eta_{8},\pi^+\pi^-} = G_{\eta_{8},\pi^0\pi^0}$ and
$G_{\eta_{1},\pi^+\pi^-} = G_{\eta_{1},\pi^0\pi^0}$. We observe
then from Eqs.~(\ref{five}) that in this limit we obtain the
so-called $|\Delta I|=1/2$ rule predictions: $A_{+0} = 0$ and
$A_{+-}=A_{00}$. The so-called parity-violating amplitude
$A_{+0}$ vanishes already in the SU(2) symmetry limit.

In the SU(3) symmetry limit, where $G_{\eta_{8},\pi^0\pi^0} =
-(1/\sqrt{3})G_{\pi^0,\pi^0\pi^0}$, the expressions for the
remaining so-called parity-violating amplitudes $A_{+-}$ and
$A_{00}$ take the form

\begin{equation}
A_{+-}=A_{00}=
[-\frac{1}{3}(2\delta + \delta') G^{\rm
s,pp}_{\pi^0,\pi^0\pi^0} +
\sqrt{\frac{2}{3}}(\delta-\delta')G^{\rm
s,pp}_{\eta_1,\pi^0\pi^0}]e^{i\alpha_{1}}.
\label{fivesu3}
\end{equation}

\noindent
In this case, the vanishing of $A_{+-}$ and $A_{00}$ requires not
only that $\delta'=\delta$ of conditions~(\ref{cond}) or
(\ref{condwrd}) be satisfied but it also requires the use of the
generalized Bose principle~\cite{neubert} which becomes
operative in the SU(3) symmetry limit and forces the matrix
elements $\langle\pi\pi |H_{\rm st}| \pi\rangle$ to vanish, as
described above.

\section{Conclusions}
\label{conclusions}

In this paper we have extended our previous analysis on the SU(3)
symmetry limit properties of the so-called parity-conserving
amplitudes of NLDH and WRDH~\cite{symmetry} to all the amplitudes
involved in these two groups, in the two-body non-leptonic
decays of $\Omega^-$, and in the two photon and two-body
non-leptonic decays of $K$'s.

The amplitudes have been obtained by direct application
of expressions~(\ref{mph}) and (\ref{bph}) for the physical hadrons
with mirror matter admixtures. The contributions of these mixings
were determined by the matrix elements of the parity and
strong-flavor conserving strong and electromagnetic parts of the
exact Hamiltonian.

The so-called parity-conserving amplitudes vanish automatically
in the strong flavor SU(3) symmetry limit. So that we may
restate our conclusion by saying that even if mixings with parity
conserving but flavor violating eigenstates are allowed in
physical hadrons (i.e., $\sigma\ne 0$) the mixing angle $\sigma$
will drop out of the matrix elements of the strong and
electromagnetic parts of the Hamiltonian that contribute to
non-leptonic, weak radiative, and two photon decays
of strange hadrons in the strong flavor SU(3) symmetry limit.

The so-called parity-violating amplitudes vanish if
$\delta'=\delta$ and a relative minus sign be present in the
Yukawa couplings and transitions magnetic moments as shown in
Eqs.~(\ref{cond}) and (\ref{condwrd}). The vanishing of the $K_{\rm
S}\to\gamma\gamma$ amplitude requires only that $\delta'=\delta$.
The vanishing of the so-called parity-violating $K\to\pi\pi$
amplitudes also requires $\delta'=\delta$ and the extended
Bose-Einstein statistics to the $s=0$ meson octet in the SU(3)
symmetry limit. We restate our conclusion by saying that even if
parity and flavor violating admixtures were allowed in physical
hadrons the mixing angle $\delta'=\delta\ne 0$ will drop out of
the matrix elements of the strong and electromagnetic parts of
the Hamiltonian that contribute to the above decays in the
strong flavor SU(3) symmetry limit.

Notice that the decays $K_{\rm L,S}\to\mu+\mu-$ are
automatically covered due to the central role that $K_{\rm
L,S}\to\gamma\gamma$ play in these decays, respectively.

One should contrast our previous results on the so-called
parity-conserving amplitudes and our new results on the
so-called parity-violating amplitudes with the
existing theorems for the $W$-mediated non-leptonic and weak
radiative decays of hyperons, which are referred to as the
Lee-Swift~\cite{ref5} and Hara~\cite{ref6} theorems,
respectively. Both these theorems state that in the flavor
symmetry limit it is the corresponding parity-violating
amplitudes that vanish. The theorem of Ref.~\cite{ref5} refers
only to the baryon pole contributions to the parity-violating
amplitudes of non-leptonic decays of hyperons. The theorem of
Ref.~\cite{ref6} is limited to $\Sigma^+\to p\gamma$ and
$\Xi^-\to\Sigma^-\gamma$. Our results here cover more cases than
these two theorems and in this sense are more general.

\acknowledgments

We would like to thank CONACyT (M\'exico) for partial support.

\end{document}